# To Apply or Not to Apply:
# A Survey Analysis of Grant Writing Costs and Benefits

Ted von Hippel
Embry-Riddle Aeronautical University
Daytona Beach, FL 32114, USA
Corresponding author: ted.vonhippel@erau.edu

Courtney von Hippel
University of Queensland
St Lucia, QLD, 4069, Australia

**Abstract**

We surveyed 113 astronomers and 82 psychologists active in applying for federally funded research on their grant-writing history between January, 2009 and November, 2012. We collected demographic data, effort levels, success rates, and perceived non-financial benefits from writing grant proposals. We find that the average proposal takes 116 PI hours and 55 CI hours to write; although time spent writing was not related to whether the grant was funded. Effort did translate into success, however, as academics who wrote more grants received more funding. Participants indicated modest non-monetary benefits from grant writing, with psychologists reporting a somewhat greater benefit overall than astronomers. These perceptions of non-financial benefits were unrelated to how many grants investigators applied for, the number of grants they received, or the amount of time they devoted to writing their proposals. We also explored the number of years an investigator can afford to apply unsuccessfully for research grants and our analyses suggest that funding rates below approximately 20%, commensurate with current NIH and NSF funding, are likely to drive at least half of the active researchers away from federally funded research. We conclude with recommendations and suggestions for individual investigators and for department heads.

## Introduction

Winning research grants from federal funding agencies in the United States is highly competitive and will likely become increasingly so [1]. For instance, agency-wide funding rates at the National Institutes of Health (NIH) and the National Science Foundation (NSF) since 2010 were 18-21% [2] and 22-24% [3], respectively. Yet evidence suggests that the fraction of meritorious proposals is likely to be much higher. For example, 68% of proposals at NSF were rated as meritorious [4] and evidence suggests a similar rate at NIH [5]. Of course, it is possible that these unfunded but positively evaluated proposals are meritorious but not excellent, and thus not truly deserving of funding in a tight budget climate in which there are many worthwhile claimants for federal dollars. In contrast to this possibility, one large-scale study revealed that the inter-rater reliability for grant application reviews is strikingly low, ranging from 0.15 to 0.2 [6]. Furthermore, a study of NIH grant recipients found no difference in scientific impact as a function of the grant proposals' percentile rankings [7]. Given the challenge inherent in evaluating the plausibility of ideas that have not yet been tested, these low reliability rates should not come as a major surprise. Nonetheless, even if we accept (hope) that reviewers can correctly identify and agree on which grants are in the top 50%, these low reliability rates suggest that being chosen for funding from the top half of all proposals represents a roll of the dice. With funding rates at 20%, that leads to a best-case scenario of 60% of these deserving proposals not being funded in any given year. For these and other reasons, there is broad agreement that federal grant agencies are underfunded [8]-[11].

With funding rates at a level such that the majority of meritorious proposals are unfunded, it is natural for researchers to ask a range of questions about the proposal process: What is the typical amount of time researchers spend writing grant proposals (i.e., what are the opportunity costs of writing proposals that are unlikely to be funded even if they are deserving)? Are there meaningful non-financial benefits to writing proposals that offset these opportunity costs? How should funding rates impact a researcher's decision of whether to write a proposal? We investigate these questions by conducting a survey of research-active astronomers and psychologists. Our goal is to provide a statistical description of key aspects of the proposal process that will be useful to researchers formulating their grant writing plans, as well as to department heads and research administrators who advise researchers on this topic.

## Methods

### Procedure and Participants

The study focused on grant proposals submitted by astronomers and social and personality psychologists to three US federal agencies that fund basic research: NASA, the NIH, and the NSF. We chose these fields because they occupy either end of a distribution, with astronomy representing a fundamental physical science and psychology representing a mixture of basic and applied social science. We advertised the survey through the American Astronomical Society, by appeals to colleagues in 24 astronomy departments, and through the list serve of the Society for Personality and Social Psychology. The survey (see S1_Survey.pdf) ran in

October and November, 2012. We obtained ethics clearance through the University of Queensland and Embry-Riddle IRB. The welcome page of the on-line survey described the purpose and ethics information of the study and participants provided their informed consent by clicking on the "begin survey" icon to enter the survey.

Participants were 195 academics from universities/colleges across the United States. The breakdown by profession was 113 astronomers (84 male; 25 female; 4 missing) and 82 psychologists (39 male; 42 female; 1 missing). The survey first asked participants if they had applied for a grant from NIH, NSF, or NASA since January 2009. If participants submitted grants to more than one of these agencies since January 2009, the survey asked "To which agency did you most recently submit a grant as a Principal/Primary Investigator?". Through survey software piping, most remaining questions in the survey focused on participants' most recent grant submission to one of these agencies.

The rank of respondents included 12 postdocs, 37 assistant professors, 34 associate professors, 59 full professors, 4 emeritus professors, 14 research assistant professors, 15 research associate professors, 6 research professors, and 8 other. Although it was not our intention to survey graduate students, 3 PhD students also completed the survey. Inclusion of their data does not impact the results and thus we report all analyses, unless otherwise noted, with the full sample of respondents. Consistent with the range of seniority, respondents earned their PhD as early as 1964 and as recently as 2012, or for the students, anticipated through 2015. There was a gentle peak of respondents who earned their PhD between 2001 and 2008.

The majority of respondents (i.e., 125) were from large research universities with PhD programs, 21 were from moderate-sized research active universities with some PhD programs, 8 from undergraduate-focused colleges/universities with PhD programs, 14 from undergraduate-focused colleges/universities without PhD programs, 9 from government departments or labs, 15 from a type of institution not specified, and 3 people who did not answer the question.

**Measures**

***Grant history***. To establish the history of participants' latest grant application, the survey asked researchers what year their most recent submission was (between January 2009 and the survey period) and the amount of funding they had requested. The survey then asked participants if the grant proposal received funding (with response options *yes*, *no*, *not yet known*, and *the project was rated as fundable, pending budget approval*).

***Effort expenditure.*** The survey asked participants to provide an estimate of the total number of hours they spent preparing the grant application. The survey indicated that this estimate should include background reading, data analyses, writing the proposal, preparing the budget, generating letters of support from departmental heads or other administrators, reading the funding agency documentation, etc. The survey further explained that the estimate should include any grant related activity. The survey also asked participants to provide an estimate of the number of hours all other investigators (combined) spent working on the grant application.

***Non-financial benefits of grant writing.*** To explore the degree to which researchers experienced non-financial benefits of writing grant proposals, the survey asked participants to indicate how much they agreed with nine statements regarding potential benefits ($\alpha$ = .87) on a 7-point scale anchored by *strongly disagree* (1) and *strongly agree* (7). These items were: "Writing a grant…advances or fine tunes my scientific thinking; enables me to consolidate or organize my research efforts/plans; helps me generate new ideas that I wouldn't have had otherwise; helps me plan the workflow for my research group; helps train/educate my graduate students and/or postdocs; helps me develop new collaborations; helps me focus on the big picture rather than just the details of my projects; results in text that I can then use for future papers and/or conference submissions; and, for me there are no benefits to grant writing except getting the grant" (reverse scored).

***Grant history outcomes.*** To examine grant writing history, the survey asked participants how many grants they submitted to each granting agency (i.e., NASA, NIH, and NSF) over a 3-year period (i.e., 2009-2011). Participants then indicated how many of these applications were funded.

***Demographics.*** Participants responded to questions assessing their age, gender, current salary, current position (e.g., assistant professor, associate professor, etc.), type of institution (e.g., large university with many PhD students), number of years employed at their current institution, and the year in which their PhD was awarded.

## Results

We used the statistical software package SPSS to analyze the data (available in S1_Dataset.csv). The vast majority of participants from psychology were in tenure-track or tenured positions (70 of 74 academics, in addition to the 8 participants who were PhD students, post-docs, or emeritus professors). In contrast, a substantial minority from astronomy was in research-only positions (60 tenure-track or tenured; 31 research only, and 22 participants who were PhD students, post-docs, or emeritus professors). To determine if research-only astronomers differed in their grant activities and perceived non-financial benefits of grant writing from tenure-track/tenured faculty in astronomy, we examined these two groups separately. Our analyses revealed no significant differences between tenure-track/tenured and research-only faculty (all $F$'s ≤ 1.32; all $p$'s > .25) and thus we report the results across all academic positions combined.

Table 1 presents the means and standard deviations of the measures. On average PIs spent 116 hours and CIs spent 55 hours per proposal. To put these numbers in perspective, data from a survey [12] of 23,824 full-time faculty at 417 four-year colleges and universities reveal that faculty at public universities, private universities, public four-year colleges, and private four-year colleges spend an average of 10.7, 14.5, 5.6, and 5.7 hours per week on research and scholarly writing during the teaching semesters, respectively. Seventy-six percent of our respondents were employed at large research universities, corresponding to the 10.7 to 14.5 hour-per-week values.  For an individual corresponding to these averages, a proposal requires at least 8.0 PI weeks and 3.8 CI weeks.  While successful PIs and CIs may have more time per week for research than these averages, and research-only academics assuredly have more time for research than teaching academics, 116 PI and 55 CI hours likely represents a minimum commitment of one month of full-time effort.

Table 1 also presents the inter-correlations of the measures.  The time a PI invests in writing a given proposal was positively correlated with the hours spent by the remainder of the proposing team, yet reported hours of effort was not related to whether the grant was funded.  Perhaps reassuringly, academics who wrote more grants received more funding, although the causal direction could go either way.  In terms of differences between the disciplines, psychologists submitted fewer grants but they spent more hours on each submission and found the grant writing process more rewarding than astronomers.  Not surprisingly, the relationships between tenure, salary, and rank was such that tenured academics enjoying higher salaries and higher ranks.  The PI's gender was correlated with tenure status and salary, reflecting the well-known decreasing ratio of women at higher salaries and in more senior positions in academics [13], [14].  Gender was also related to area of study, with a higher ratio of female academics among psychologists than astronomers.  Despite the correlation between gender and salary, it does not appear that grant activity drives this relationship, in that gender was not correlated with number of grants applied for or won or the dollar amount requested.

**Measures**

***Grant history***.  Not surprisingly given the topic of the survey, respondents were typically active in writing research proposals.  Participants submitted 287 NASA, 124 NIH, and 345 NSF grant proposals between January, 2009 and the date on which they participated in the survey, or an average of 3.88 proposals per respondent over 3.75 years.  Their most recent grant applications were distributed as follows: 63 to NASA (astronomers), 42 to NIH (psychologists), and 90 to NSF (50 from astronomers; 40 from psychologists).

***Non-financial benefits of grant writing***.  Table 2 provides the mean and 95% confidence intervals for each individual item, separately for psychology and astronomy participants.  Participants perceived various non-monetary benefits from grant writing, with psychologists reporting somewhat greater benefits ($M$ = 5.21; $SD$ = .87) overall than astronomers ($M$ = 4.66; $SD$ = 1.18), $F(1, 192) = 12.64$, $p < .01$.  As can be seen in Table 2, participants more strongly endorsed some benefits than others.  For example, there is stronger support for the notion that grant writing benefits their scientific thinking than for the idea that it helps them develop collaborations, train students and postdocs, or manage lab workflow.  Nevertheless, there was no relationship between perceptions of such non-financial benefits and how many grants investigators applied for or received, nor was there a relationship between perceived non-financial benefits and amount of time investigators devoted to writing their proposals (see Table 1).

**Comparing survey participants to members of their field**

Table 3 summarizes the success rates of survey participants and compares these data to the agency funding rates. The columns cover the four combinations of academic field/funding agency and the rows correspond to (1) the proposal success rate of participants in our survey and (2) the agency success rates for each of 2009 through 2012 in those fields. The success rate for astronomers in this sample is representative of the field as a whole, whereas the success rate for psychologists is somewhat higher than typical for psychologists applying to these funding agencies.

## Discussion

We proposed at the outset of this paper that unfunded science is not necessarily poor science. Because a large proportion of meritorious science is not being funded, consistently failing to obtain research support is a realistic prospect for many excellent researchers. As noted in a 2008 AAAS report [10], "One-half of [NSF] new investigators never again receive NSF funding after their initial award." Some of our survey respondents remarked on this explicitly, writing "I applied for grants from the NSF in 2004, 2005, 2006 and 2007. … Most of the reasons given for not funding were that funds were too tight that particular year and that I should reapply the next year since the proposal had merit . . . I finally just gave up." Or "I ceased to apply for grants as a PI after 2007 when – after much experience – I decided that applying for federal grants was not a good use of my time". A research-active department head wrote, "I don't feel that it is worth my time to apply." In these ways, the experiences of our sample are consistent with national statistics.

How does the funding rate connect to this period of applying but failing to obtain grants? We begin with the assumption, to be refined below, that the probability of writing a successful grant is independent of grant successes that an investigator has had in the past. This assumption is motivated by the fact that funding rates are low and the inter-rater reliability for grant application reviews is inadequate [6], resulting in funding decisions being based to a substantial degree on chance. Assuming independence in funding probabilities from one proposal to the next, the chance of failing to obtain any grants after n attempts is $(1 - \text{funding rate})^n$. For investigators taking advantage of two opportunities per year and a 20% funding rate, which has been characteristic of psychology over the last decade and characteristic of astronomy over the last few years, the probability of failing to obtain any funding after three years of effort is 26%. Yet, writing two grant proposals per year for multiple years is a substantial burden, requiring at least one month of dedicated PI and CI research investment per proposal. It is therefore not surprising that respondents wrote an average of one grant proposal per year. Writing one proposal per year to agencies with funding rates of 20% would lead to 51% of applicants receiving no funding after three years of effort.

The above formula has the advantage of simplicity, which allows individual investigators to quickly personalize it for a different funding rate or planned number of attempts. Yet, the assumption of independence is problematic. We therefore turn to our survey to examine this assumption. Fifty researchers within our survey applied for NASA, NIH, or NSF grants during the period 2009-11, applied to these same agencies during 2012, and knew the agency decision on their proposal when they completed the survey in late 2012. We used their data to calculate the conditional probabilities of obtaining funding in 2012 given that funding was or was not obtained in the period 2009-11, P(present funding | past funding) and P(present funding | no past funding). We find P(present funding | past funding) = 17 out of 35 proposers ~ 50% and P(present funding | no past funding) = 1 out of 15 proposers ~7%.

Although the sample size is small, it is clear that researchers who have enjoyed recent proposal success face substantially better odds than the current base rate and researchers who have not had recent success face substantially worse odds (known as the Matthew Effect [15] or Cumulative Advantage). For those researchers with past grant success, the relatively high value of P(present funding | past funding) appears to be good news, yet it also means that low funding rates are even lower than they appear for investigators who have not had recent success (more on this below). Indeed, even among highly capable researchers with productive teams, 50% were unable to obtain funding in the current cycle. After three such cycles, one-eighth of all active programs are likely to be defunded.

Additionally, if presently funded teams obtain one half of the available funding in the current cycle, and with a funding rate of 20%, presently unfunded and new investigators are competing with one another for an effective funding rate of only ~12% (80% of proposers competing for 10% of the funds). The assumption of independence across grant proposal success probabilities is thus optimistic for all but a small minority of applicants. In fact, even among investigators for whom the assumption of independence is valid (e.g., new investigators without a substantial grant-writing track record or more experienced researchers who have not recently sought funding), such individuals are not competing for the true funding rate, but rather for a lower effective funding rate. A 12% effective funding rate for this group, for example, would mean that 68% would not receive funding after three attempts. Using the conditional probabilities found in our survey, one-eighth of the presently funded researchers (0.12 x 0.2) plus two-thirds of the presently unfunded or new researchers (0.68 x 0.8), or a total of 78% of proposers, will be unable to secure federal funds for their research.

This rate is substantially higher than the ~50% driven from federally funded research derived above, because the independence calculation assumes everyone is on an equal footing and there will be a high rate of turn-over, whereas the conditional probability calculation indicates that a subset of researchers are able to hold onto funding, making less funding available for new investigators or those who wrote previously unsuccessful proposals. Therefore, the reality for the majority of those writing grant proposals is worse than the independence calculation implies and the results of that calculation are conservative. For these reasons, we conservatively estimate that funding rates of 20% will force one half of grant applicants to abandon federally funded research after a multi-year effort. Not surprisingly, failing to obtain grants after years of steady effort is often the end of that line of research for most investigators [16], [17]. If these lines of unfunded research were of relatively low quality this outcome would not necessarily be problematic, but our earlier analysis suggests that much of this research is indistinguishable in quality from research that does receive funding.

## Recommendations

Because a 20% funding rate will force at least half of all proposers to abandon federally funded research after multiple years of effort, we recommend that proposers, research mentors, and funding agencies compare current funding rates to this value. We suggest that individual investigators should consider avoiding proposing to programs with funding rates at or below 20% unless they are confident that their research program has a greater-than-baseline chance of success or they are willing to write two or more proposals per year. If researchers have a recent history of writing successful or unsuccessful proposals to this grant agency, we suggest that they consider our discussion of conditional probabilities as it pertains to their situation. We recommend that department heads and research administrators think carefully about which researchers they should guide toward low funding rate programs, basing their decisions on realistic chances of success and time available for writing proposals. We note that 20% funding rates impose a substantial opportunity cost on researchers by wasting a large fraction of the available research time for at least half of our scientists, reducing national scientific output, and driving many capable scientists away from productive and potentially valuable lines of research.

**Acknowledgements:** We acknowledge valuable feedback that helped refine our manuscript from Drs. E. J. Mierkiewicz, T. D. Oswalt, A. Sarajedini, and W. von Hippel.